\documentclass[supp]{new_tlp} %%added supp option
\usepackage[utf8]{inputenc}
\usepackage[T1]{fontenc} 
\usepackage{graphicx}
\usepackage{wrapfig}
\usepackage{supp} %%added

\begin{document}

\title[A Simple and Efficient Lock-Free Hash Trie Design for Concurrent Tabling]
      {A Simple and Efficient Lock-Free Hash Trie Design\\for Concurrent Tabling}

\author[M. Areias and R. Rocha]
       {MIGUEL AREIAS and RICARDO ROCHA\\
       CRACS \& INESC TEC, Faculty of Sciences, University of Porto\\
       Rua do Campo Alegre, 1021/1055, 4169-007 Porto, Portugal\\
       \email{\{miguel-areias,ricroc\}@dcc.fc.up.pt}
      }

\maketitle

%%%%%%%%%%%%%%%%%%%%%%%%%%%%%%%%%%%%%%%%%%%%%%%%%%%%%%%%%%%%%%%%%%%%%%%%%%%%

\begin{abstract}
  A critical component in the implementation of a concurrent tabling
  system is the design of the table space. One of the most successful
  proposals for representing tables is based on a \emph{two-level trie
    data structure}, where one trie level stores the tabled subgoal
  calls and the other stores the computed answers. In this work, we
  present a simple and efficient \emph{lock-free design} where both
  levels of the tries can be shared among threads in a concurrent
  environment. To implement lock-freedom we took advantage of the
  \emph{CAS} atomic instruction that nowadays can be widely found on
  many common architectures. \emph{CAS} reduces the granularity of the
  synchronization when threads access concurrent areas, but still
  suffers from low-level problems such as false sharing or cache
  memory side-effects. In order to be as effective as possible in the
  concurrent search and insert operations over the table space data
  structures, we based our design on a \emph{hash trie data structure}
  in such a way that it minimizes potential low-level synchronization
  problems by dispersing as much as possible the concurrent
  areas. Experimental results in the Yap Prolog system show that our
  new lock-free hash trie design can effectively reduce the execution
  time and scale better than previous designs.
\end{abstract}


%%%%%%%%%%%%%%%%%%%%%%%%%%%%%%%%%%%%%%%%%%%%%%%%%%%%%%%%%%%%%%%%%%%%%%%%%%%%

\section{Introduction}

Tabling~\cite{Chen-96} is a recognized and powerful implementation
technique that overcomes some limitations of traditional Prolog
systems in dealing with recursion and redundant
sub-computations. Multithreading in Prolog is the ability to perform
concurrent computations, in which each thread runs independently but
shares the program clauses~\cite{Moura-08b}. Despite the availability
of both multithreading and tabling in some Prolog systems, the
efficient implementation of these two features, such that they work
together, implies a complex redesign of several components of the
underlying engine. XSB was the first Prolog system to combine tabling
with multithreading~\cite{Marques-08}. In more recent
work~\cite{Areias-12a}, we have proposed an alternative view to XSB's
approach, where each thread views its tables as private but, at the
engine level, we use a \emph{common table space}, i.e., from the
thread point of view, the tables are private but, from the
implementation point of view, tables are shared among all threads.

A critical component in the implementation of an efficient tabling
system is the design of the data structures and algorithms to access
and manipulate tabled data. To deal with concurrent table accesses,
our initial approach, implemented on top of the Yap Prolog
system~\cite{CostaVS-12}, was to use \emph{lock-based} data
structures~\cite{Areias-12a}. Yap implements the table space using a
two-level trie data structure, where one trie level stores the tabled
subgoal calls and the other stores the computed answers. More
recently~\cite{Areias-14}, we presented a sophisticated
\emph{lock-free} design to deal with concurrency in both trie
levels. Lock-freedom allows individual threads to starve but
guarantees system-wide throughput. To implement lock-freedom we took
advantage of the $CAS$ atomic instruction that nowadays can be widely
found on many common architectures. The $CAS$ reduces the granularity
of the synchronization when threads access concurrent areas, but still
suffers from contention points where synchronized operations are done
on the same memory locations, leading to low-level problems such as
false sharing or cache memory ping pong side-effects.

In this work, we go one step further and we present a simpler and
efficient lock-free design based on \emph{hash tries} that minimizes
these problems by dispersing as much as possible the concurrent
areas. Hash tries (or hash array mapped tries) are a trie-based data
structure with nearly ideal characteristics for the implementation of
hash tables~\cite{Bagwell-01}. An essential property of the trie data
structure is that common prefixes are stored only
once~\cite{Fredkin-62}, which in the context of hash tables allows us
to efficiently solve the problems of setting the size of the initial
hash table and of dynamically resizing it in order to deal with hash
collisions. The aim of our proposal is to be as effective as possible
in the search and insert operations, by exploiting the full
potentiality of lock-freedom on those operations, and in such a way
that it minimizes the bottlenecks and performance problems mentioned
above without introducing significant overheads for sequential
execution.

Several approaches do exist in the literature for the implementation
of lock-free hash tables, such as Shalev and Shavit split-ordered
lists~\cite{Shalev-06}, Triplett \emph{et al.}  relativistic hash
tables~\cite{Triplett-11} or Prokopec \emph{et al.}
CTries~\cite{Prokopec-12}. However, to the best of our knowledge, none
of them is specifically aimed for an environment with the
characteristics of our tabling framework that does not requires
concurrent deletion support. In general, a tabled program is
deterministic, finite and only executes search and insert operations
over the table space data structures. In Yap Prolog, space is
recovered when the last running thread abolish a table. Since no
delete operations are performed, the size of the tables always grows
monotonically during an evaluation. Initial experiments, on top of a
32 core AMD machine, show that our new lock-free hash-trie design can
effectively reduce the execution time and scale better than all the
previously implemented lock-based and lock-free strategies.

%%%%%%%%%%%%%%%%%%%%%%%%%%%%%%%%%%%%%%%%%%%%%%%%%%%%%%%%%%%%%%%%%%%%%%

\section{Background}

\begin{wrapfigure}{r}{2.4cm}
\vspace{-\intextsep}
\centering
\includegraphics[width=2cm]{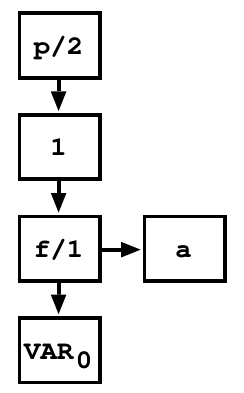}
\caption{Trie example}
\label{fig_trie_example}
\vspace{-\intextsep}
\end{wrapfigure}

A trie is a tree structure where each different path corresponds to a
term described by the tokens labeling the nodes traversed. For
example, the tokenized form of the term $p(1,f(X))$ is the sequence of
4 tokens $p/2$, $1$, $f/1$ and $VAR_0$, where each variable is
represented as a distinct $VAR_i$ constant. Two terms with common
prefixes will branch off from each other at the first distinguishing
token. Consider, for example, a second term $p(1,a)$. Since the main
functor and the first argument, tokens $p/2$ and $1$, are common to
both terms, only one additional node will be required to fully
represent this second term in the trie. Figure~\ref{fig_trie_example}
shows Yap's trie structure that represents both terms.

Whenever the chain of child nodes for a common parent node becomes
larger than a predefined threshold value, a \emph{hash mechanism} is
used to provide direct node access and therefore optimize the
search. To deal with hash collisions, all previous Yap's approaches
implemented a dynamic resizing of the hash tables by doubling the size
of the bucket entries in the hash. Our initial approach to support
concurrent tabling was \emph{lock-based}, which required
synchronization between threads when performing the hash expansion
procedure~\cite{Areias-12a}. More recently, we proposed a
\emph{lock-free} design for concurrent table accesses that avoids
thread synchronization, even when threads are expanding the hash
tables~\cite{Areias-14}. In this work, we present a simpler and
efficient lock-free design based on \emph{hash tries} to implement the
hash mechanism inside the subgoal and answer tries.

\begin{wrapfigure}{r}{7cm}
\vspace{-\intextsep}
\centering
\includegraphics[width=7cm]{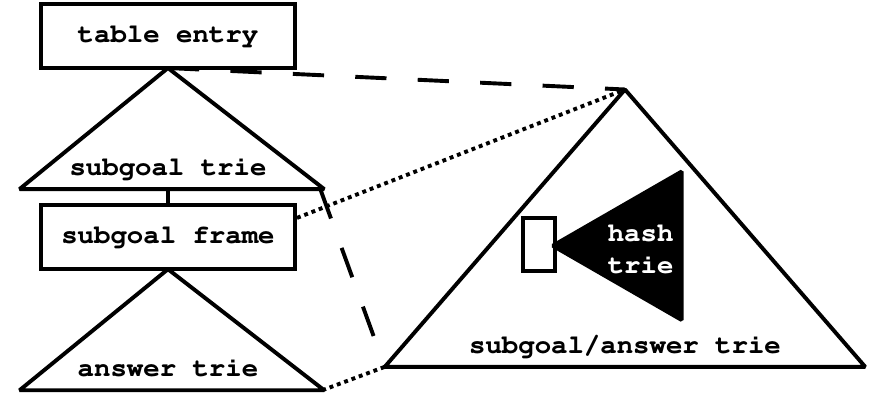}
\caption{Trie hierarchical levels overview}
\label{fig_trie_levels}
\vspace{-\intextsep}
\end{wrapfigure}

To put our proposal in perspective, Fig.~\ref{fig_trie_levels} shows a
schematic representation of the trie hierarchical levels we are
proposing to implement Yap's table space. For each predicate being
tabled, Yap implements tables using two levels of tries together with
the \emph{table entry} and \emph{subgoal frame} auxiliary data
structures~\cite{Rocha-05a}. The first level, the \emph{subgoal trie},
stores the tabled subgoal calls and the second level, the \emph{answer
  trie}, stores the answers for a given call. Then, for each
particular subgoal/answer trie, we have as many trie levels as the
number of parent/child relationships (for example, the trie in
Fig.~\ref{fig_trie_example} has 4 trie levels). Finally, to implement
hashing inside the subgoal/answer tries, we use another trie-based
data structure, the hash trie, which is the focus of the current
work. In a nutshell, a hash trie is composed by \emph{internal hash
  arrays} and \emph{leaf nodes}. The leaf nodes store $key$ values and
the internal hash arrays implement a hierarchy of hash levels of fixed
size $2^w$. To map a $key$ into this hierarchy, we first compute the
hash value $h$ for $key$ and then use chunks of $w$ bits from $h$ to
index the entry in the appropriate hash level. Hash collisions are
solved by simply walking down the tree as we consume successive chunks
of $w$ bits from the hash value $h$.

%%%%%%%%%%%%%%%%%%%%%%%%%%%%%%%%%%%%%%%%%%%%%%%%%%%%%%%%%%%%%%%%%%%%%%

\section{Our Proposal By Example}

\begin{figure}[ht]
\centering
\includegraphics[width=10.25cm]{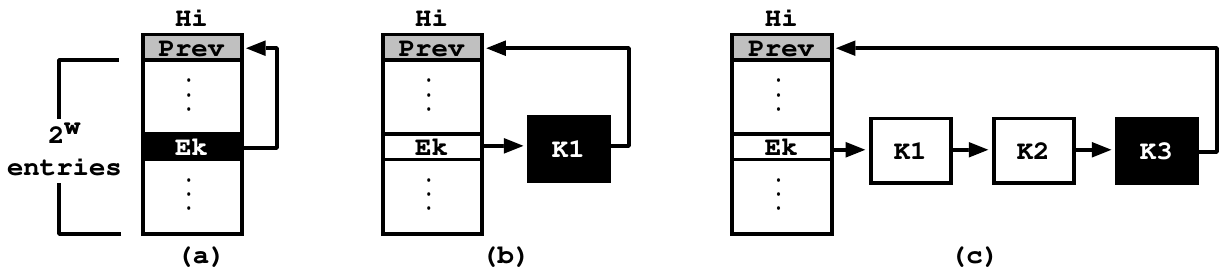}
\caption{Insert procedure in a hash level}
\label{fig_first_level}
\end{figure}

We will use three examples to illustrate the different configurations
that the hash trie assumes for one, two and three levels (for more
levels, the same idea applies). We begin with
Fig.~\ref{fig_first_level} showing a small example that illustrates
how the concurrent insertion of nodes is done in a hash level.

Figure~\ref{fig_first_level}(a) shows the initial configuration for a
hash level. Each hash level $H_i$ is formed by a bucket array of $2^w$
entries and by a backward reference to the previous level (represented
as $Prev$ in the figures that follow). For the root level, the
backward reference is $Nil$. In Fig.~\ref{fig_first_level}(a), $E_k$
represents a particular bucket entry of the hash level. $E_k$ and the
remaining entries are all initialized with a reference to the current
level $H_i$. During execution, each bucket entry stores either a
reference to a hash level or a reference to a separate chaining
mechanism, using a chain of internal nodes, that deals with the hash
collisions for that entry. Each internal node holds a $key$ value and
a reference to the next-on-chain internal
node. Figure~\ref{fig_first_level}(b) shows the hash configuration
after the insertion of node $K_1$ on the bucket entry $E_k$ and
Fig.~\ref{fig_first_level}(c) shows the hash configuration after the
insertion of nodes $K_2$ and $K_3$ also in $E_k$. Note that the
insertion of new nodes is done at the end of the chain and that any
new node being inserted closes the chain by referencing back the
current level.

During execution, the different memory locations that form a hash trie
are considered to be in one of the following states: \emph{black},
\emph{white} or \emph{gray}. A black state represents a memory
location that can be updated by any thread (concurrently). A white
state represents a memory location that can be updated only by one
(specific) thread (not concurrently). A gray state represents a memory
location used only for reading purposes. As the hash trie evolves
during time, a memory location can change between black and white
states until reaching the gray state, where it is no further
updated.

The initial state for $E_k$ is black, because it represents the next
synchronization point for the insertion of new nodes. After the
insertion of node $K_1$, $E_k$ moves to the white state and $K_1$
becomes the next synchronization point for the insertion of new
nodes. To guarantee the property of lock-freedom, all updates to black
states are done using CAS operations. Since we are using single word
$CAS$ operations, when inserting a new node in the chain, first we set
the node with the reference to the current level and only then the
$CAS$ operation is executed to insert the new node in the chain.

\begin{figure}[ht]
\centering
\includegraphics[width=\textwidth]{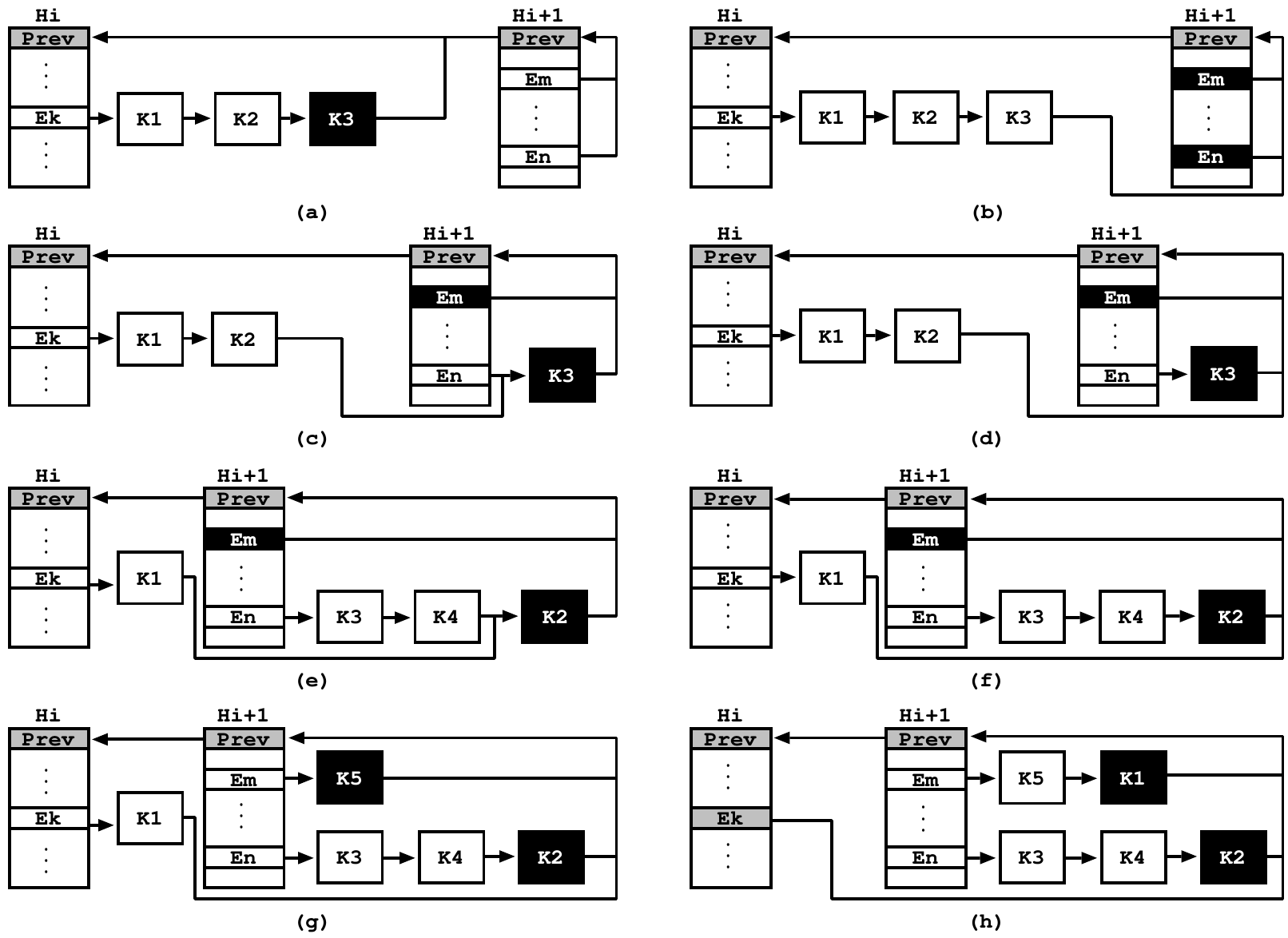}
\caption{Expanding a bucket entry with a second level hash}
\label{fig_second_level}
\end{figure}

When the number of nodes in a chain exceeds a $MAX\_NODES$ threshold
value, then the corresponding bucket entry is expanded with a new hash
level and the nodes in the chain are remapped in the new level. Thus,
instead of growing a single monolithic hash table, the hash trie
settles for a hierarchy of small hash tables of fixed size $2^w$. To
map our key values into this hierarchy, we use chunks of $w$ bits from
the hash values computed by our hash function. For example, consider a
$key$ value and the corresponding hash value $h$. For each hash level
$H_i$, we use the $w*i$ least significant bits of $h$ to index the
entry in the appropriate bucket array, i.e., we consume $h$ one chunk
at a time as we walk down the hash levels. Starting from the
configuration in Fig.~\ref{fig_first_level}(c),
Fig.~\ref{fig_second_level} illustrates the expansion mechanism with a
second level hash $H_{i+1}$ for the bucket entry $E_k$.

The expansion procedure is activated whenever a thread $T$ meets the
following two conditions: (i) the key at hand was not found in the
chain and (ii) the number of nodes in the chain is equal to the
threshold value (in what follows, we consider a threshold value of
three nodes). In such case, $T$ starts by pre-allocating a second
level hash $H_{i+1}$, with all entries referring the respective level
(Fig.~\ref{fig_second_level}(a)). At this stage, the bucket entries in
$H_{i+1}$ can be considered white memory locations, because the hash
level is still not visible for the other threads. The new hash level
is then used to implement a synchronization point with the last node
on the chain (node $K_3$ in the figure) that will correspond to a
successful $CAS$ operation trying to update $H_i$ to $H_{i+1}$
(Fig.~\ref{fig_second_level}(b)). From this point on, the insertion of
new nodes on $E_k$ will be done starting from the new hash level
$H_{i+1}$.

If the $CAS$ operation fails, that means that another thread has
gained access to the expansion procedure and, in such case, $T$ aborts
its expansion procedure. Otherwise, $T$ starts the remapping process
of placing the internal nodes $K_1$, $K_2$ and $K_3$ in the correct
bucket entries in the new level. Figures~\ref{fig_second_level}(c)
to~\ref{fig_second_level}(h) show the remapping sequence in
detail. For simplicity of illustration, we will consider only the
entries $E_m$ and $E_n$ on level $H_{i+1}$ and assume that $K_1$,
$K_2$ and $K_3$ will be remapped to entries $E_m$, $E_n$ and $E_n$,
respectively. In order to ensure lock-free synchronization, we need to
guarantee that, at any time, all threads are able to read all the
available nodes and insert new nodes without any delay from the
remapping process. To guarantee both properties, the remapping process
is thus done in reverse order, starting from the last node on the
chain, initially $K_3$.

Figure~\ref{fig_second_level}(c) then shows the hash trie
configuration after the successful $CAS$ operation that adjusted node
$K_3$ to entry $E_n$. After this step, $E_n$ moves to the white state
and $K_3$ becomes the next synchronization point for the insertion of
new nodes on $E_n$. Note that the initial chain for $E_k$ has not been
affected yet, since $K_2$ still refers to $K_3$. Next, on
Fig.~\ref{fig_second_level}(d), the chain is broken and $K_2$ is
updated to refer to the second level hash $H_{i+1}$. The process then
repeats for $K_2$ (the new last node on the chain for $E_k$). First,
$K_2$ is remapped to entry $E_n$ (Fig.~\ref{fig_second_level}(e)) and
then it is removed from the original chain, meaning that the previous
node $K_1$ is updated to refer to $H_{i+1}$
(Fig.~\ref{fig_second_level}(f)). Finally, the same idea applies to
the last node $K_1$. Here, $K_1$ is also remapped to a bucket entry on
$H_{i+1}$ ($E_m$ in the figure) and then removed from the original
chain, meaning in this case that the bucket entry $E_k$ itself becomes
a reference to the second level hash $H_{i+1}$
(Fig.~\ref{fig_second_level}(h)). From now on, $E_K$ is also a gray
memory location since it will be no further updated.

Concurrently with the remapping process, other threads can be
inserting nodes in the same bucket entries for the new level. This is
shown in Fig.~\ref{fig_second_level}(e), where a new node $K_4$ is
inserted before $K_2$ in $E_n$ and, in Fig.~\ref{fig_second_level}(g),
where a node $K_5$ is inserted before $K_1$ in $E_m$. As mentioned
before, lock-freedom is ensured by the use of $CAS$ operations when
updating black state memory locations.

To ensure the correctness of the remapping process, we also need to
guarantee that the nodes being remapped are not missed by any other
thread traversing the hash trie. Please remember that any chaining of
nodes is closed by the last node referencing back the hash level for
the node. Thus, if when traversing a chain of nodes, a thread $U$ ends
in a second level hash $H_{i+1}$ different from the initial one $H_i$,
this means that $U$ has started from a bucket entry $E_k$ being
remapped, which includes the possibility that some nodes initially on
$E_k$ were not seem by $U$. To guarantee that no node is missed, $U$
simply needs to restart its traversal from $H_{i+1}$.

\begin{figure}[ht]
\centering
\includegraphics[width=9.75cm]{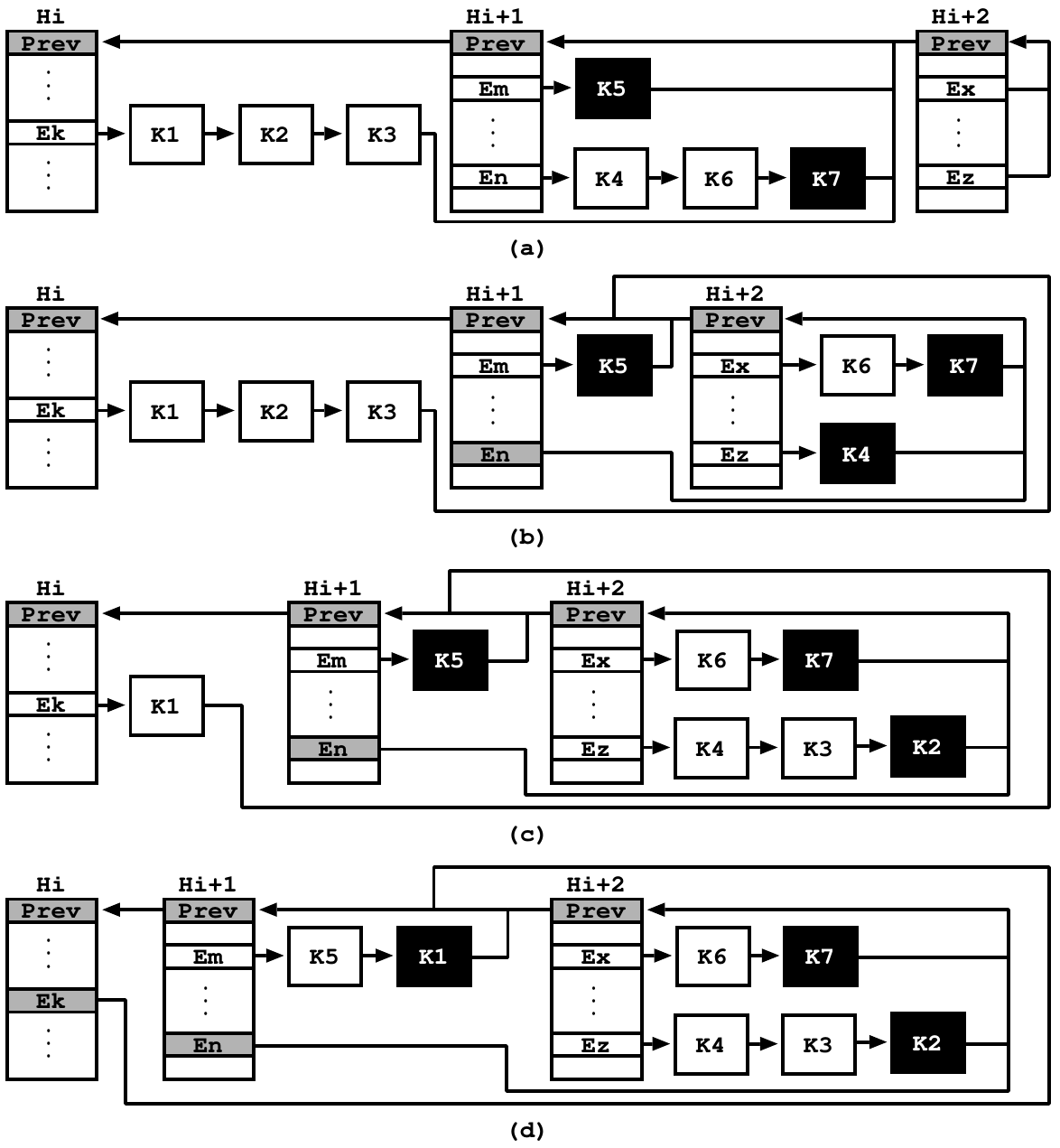}
\caption{Remapping nodes on a third level hash}
\label{fig_third_level}
\end{figure}

We conclude the description of our proposal with a last example that
shows a expansion procedure involving three hash levels. Starting from
the configuration on Fig.~\ref{fig_second_level}(b),
Fig.~\ref{fig_third_level} assumes a scenario where a set of nodes
($K_4$, $K_5$, $K_6$ and $K_7$ in the figure) are inserted in the
bucket entries $E_m$ and $E_n$ before the beginning of the remapping
process of nodes $K_1$, $K_2$ and $K_3$. Again, we will consider only
the entries $E_m$ and $E_n$ on level $H_{i+1}$ and assume that $K_1$,
$K_2$ and $K_3$ will be remapped to entries $E_m$, $E_n$ and $E_n$,
respectively.

Figure~\ref{fig_third_level}(a) shows the situation where $K_3$ is
scheduled to be remapped to entry $E_n$ on level $H_{i+1}$ but, since
the number of nodes on $E_n$ is equal to the threshold value, a
preliminary expansion procedure for $E_n$ should be done, which leads
to the pre-allocation of a third level hash
$H_{i+2}$. Figure~\ref{fig_third_level}(b) then shows the hash trie
configuration after the remapping of the nodes on $E_n$ to the level
$H_{i+2}$. Please note that $E_n$ became a gray state memory location
since it is now referring the third level hash $H_{i+2}$, which means
that any operation scheduled to $E_n$ should be rescheduled to
$H_{i+2}$. This is the case shown in Fig.~\ref{fig_third_level}(c),
where $K_3$ and $K_2$ were both rescheduled to entry $E_z$ on
$H_{i+2}$. Despite this third level remapping, the chaining reference
of the last node on the chain (for example, $K_1$ in
Fig.~\ref{fig_third_level}(c)) is still made to refer to the second
level hash $H_{i+1}$. To conclude the example,
Fig.~\ref{fig_third_level}(d) shows the configuration at the end of
the remapping process. Here, $K_1$ is remapped to the bucket entry
$E_m$ on $H_{i+1}$ and removed from the initial chain, meaning that
$E_k$ itself becomes a reference to $H_{i+1}$ and moves to a gray
state.

For each configuration shown, the reader is encourage to verify that,
at any moment, all threads are able to access all available
nodes. Consider, for example, the configuration shown in
Fig.~\ref{fig_third_level}(c) and a thread entering on level $H_i$
searching for a node with the key $K_7$. The thread would begin by
hashing the key $K_7$ on level $H_i$ and obtain the bucket entry
$E_k$. Then, it would follow the chain of nodes ($K_1$ in this case)
and reach level $H_{i+1}$. At level $H_{i+1}$, it would hash again the
key $K_7$, obtain the bucket entry $E_n$ and follow the reference to
level $H_{i+2}$. Finally, it would hash one more time the key $K_7$,
now for level $H_{i+2}$, obtain the entry $E_x$ and follow the chain
until reaching node $K_7$.

We argue that a key design decision in our approach is thus the
combination of hash tries with the use of a separate chaining (with a
threshold value) to resolve hash collisions (the original hash trie
design expands a bucket entry when a second key is mapped to
it). Also, to ensure that nodes being remapped are not missed by any
other thread traversing the hash trie, any chaining of nodes is closed
by the last node referencing back the hash level for the node, which
allows to detect the situations where a node changes level. This is
very important because it allows to implement a clean design to
resolve hash collisions by simply moving nodes between the levels. In
our design, updates and expansions of the hash levels are never done
by using replacement of data structures (i.e., create a new one to
replace the old one), which also avoids the complex mechanisms
necessary to support the recovering of the unused data
structures. Another important design decision which minimizes the
low-level synchronization problems leading to false sharing or cache
memory side-effects, is the insertion of nodes done at the end of the
separate chain. Inserting nodes at the end of the chain allows for
dispersing as much as possible the memory locations being updated
concurrently (the last node is always different) and, more
importantly, reduces the updates for the memory locations accessed
more frequently, like the bucket entries for the hash levels (each
bucket entry is at most only updated twice).

%%%%%%%%%%%%%%%%%%%%%%%%%%%%%%%%%%%%%%%%%%%%%%%%%%%%%%%%%%%%%%%%%%%%%%

\section{Performance Evaluation}

To put our results in perspective, we compared our new lock-free hash
trie design (\textbf{LFHT}) against all the previously implemented
Yap's lock-based and lock-free strategies for concurrent tabling. For
the sake of simplicity, here we will only consider Yap's best
lock-based strategy (\textbf{LB)} and the lock-free design
(\textbf{LF}) presented in~\cite{Areias-14}. For benchmarking, we used
the set of tabling benchmarks from~\cite{Areias-12b} which includes 19
different programs in total. We choose these benchmarks because they
have characteristics that cover a wide number of scenarios in terms of
trie usage. The benchmarks create different trie configurations with
lower and higher number of nodes and depths, and also have different
demands in terms of trie traversing.

Since the system's performance is highly dependent on the available
concurrency that a particular program might have, our initial goal was
to evaluate the robustness of our implementation when exposed to worst
case scenarios and, for that, we ran the benchmarks with all threads
executing the same query goal. By doing that, we avoid the
peculiarities of the program at hand and we try to focus on measuring
the real value of our new design. Since, all threads are executing the
same query goal, it is expected that all threads will access the table
space, to check/insert for subgoals and answers, at similar times,
thus stressing the synchronization on common memory locations, which
can increase the aforementioned problems of false sharing and cache
memory side-effects and thus penalize the less robust designs.

\begin{wrapfigure}{r}{4.5cm}
\vspace{-\intextsep}
\includegraphics[width=4.5cm]{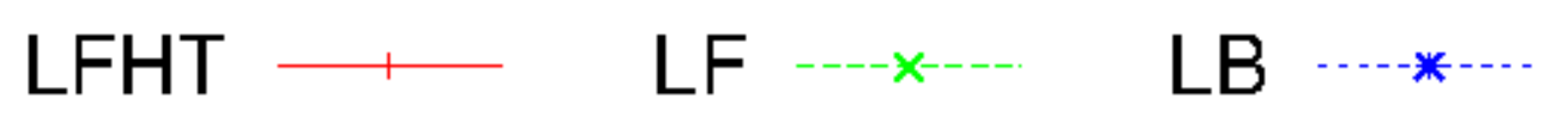}\\
\includegraphics[width=4.5cm]{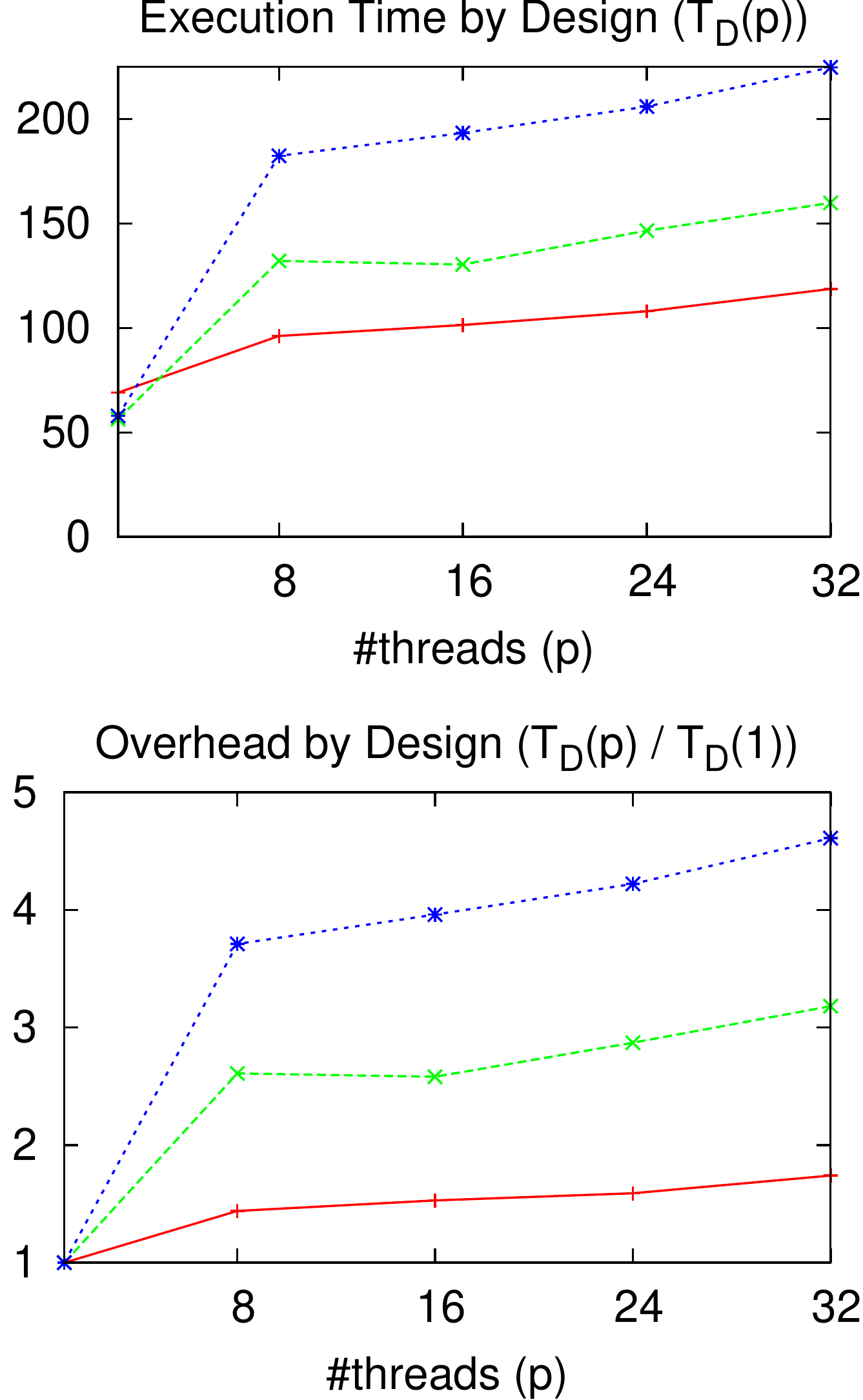}
\caption{Average execution time, in seconds, and average overhead,
  against the execution time with one thread, for the set of tabling
  benchmarks with all threads executing the same query goal}
\label{fig_eval_tab_overhead}
\vspace{-\intextsep}
\end{wrapfigure}

The environment for our experiments was a machine with 2x16 (32) Core
AMD Opteron (tm) Processor 6274 $@$ 2.2 GHz with 32 GBytes of memory
and running the Linux kernel 3.8.3-1.fc17.x86\_64 with Yap Prolog
6.3. We experimented with intervals of 8 threads up to 32 threads and
all results are the average of 5 runs for each
benchmark. Figure~\ref{fig_eval_tab_overhead} shows the average
execution time, in seconds, and the average overhead, compared against
the respective execution time with one thread, for the \textbf{LFHT},
\textbf{LF} and \textbf{LB} designs when running the set of tabling
benchmarks with all threads executing the same query goal.

The results clearly show that the new \textbf{LFHT} design achieves
the best performance for both the execution time and the overhead. As
expected, \textbf{LF} is the second best and \textbf{LB} is the
worst. In general, our design clearly outperforms the other designs
with a overhead of at most 1.74 for 32 threads (the number of cores in
the machine). Another important observation is that both \textbf{LF}
and \textbf{LB} show an initial high overhead in the execution time in
most experiments, mainly when going from 1 to 8 threads, in contrast
to \textbf{LFHT} that shows more smooth curves. The difference between
\textbf{LFHT} and \textbf{LF}/\textbf{LB} for the overhead ratio in
these benchmarks clearly shows the distinct potential of the
\textbf{LFHT} design.

\begin{wrapfigure}{r}{4.5cm}
%\vspace{-\intextsep}
\includegraphics[width=4.5cm]{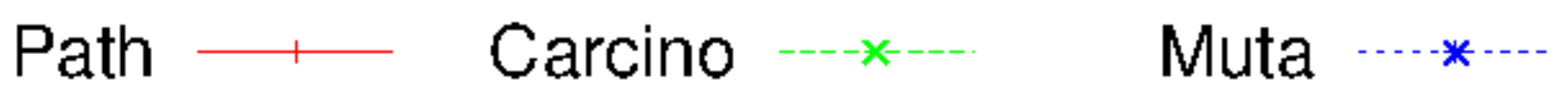}\\
\includegraphics[width=4.5cm]{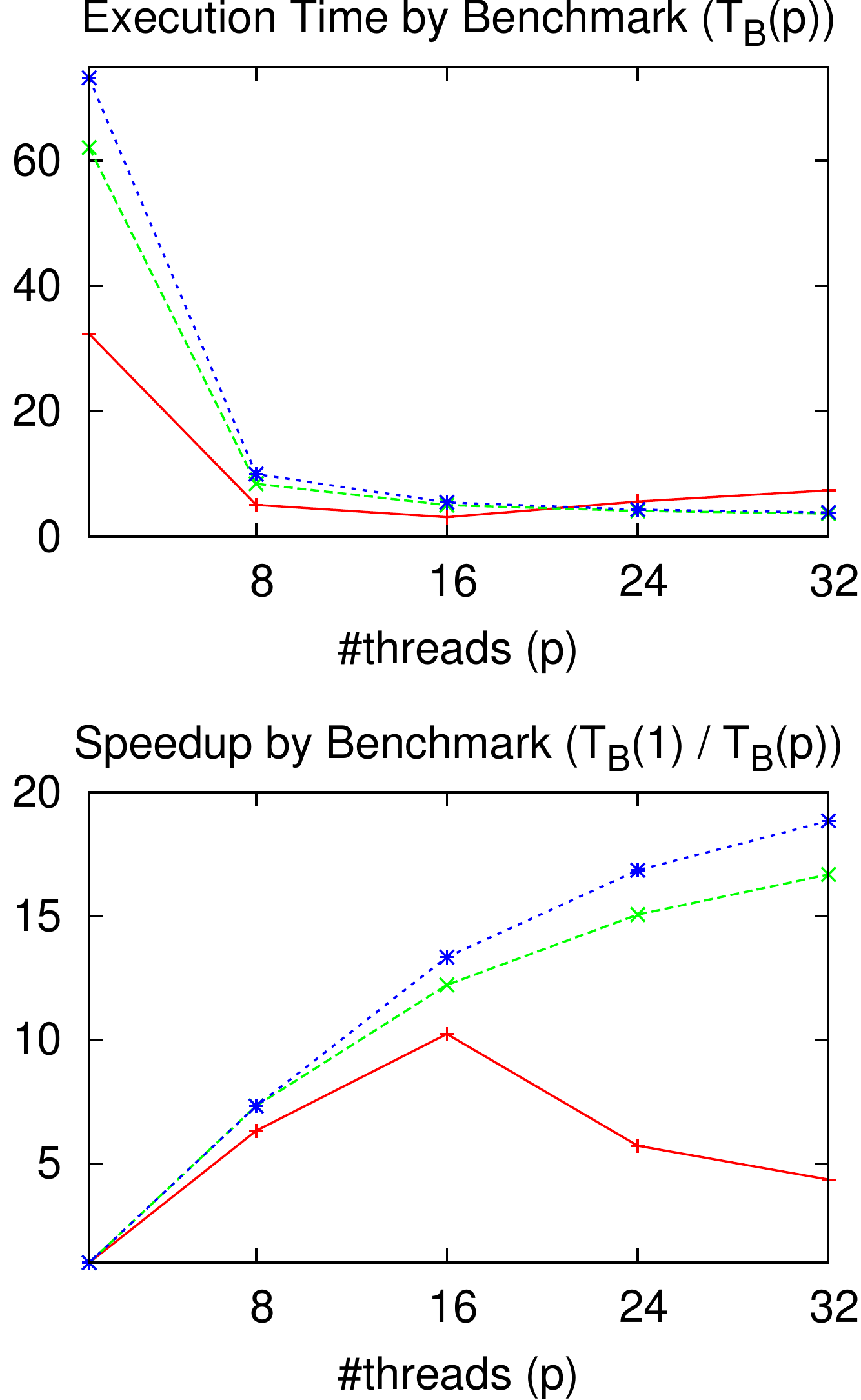}
\caption{Execution time, in seconds, and speedup, against the
  execution time with one thread, for running the naive scheduler
  program with the \textbf{LFHT} design}
\label{fig_eval_tab_speedup}
\vspace{-\intextsep}
\end{wrapfigure}

Besides measuring the value of our new design through the use of worst
case scenarios, we conclude the paper by showing the potential of our
work to speedup the execution of tabled programs. Other works have
already showed the capabilities of the use of multithreaded tabling to
speedup tabled execution~\cite{Marques-08,Marques-10}. Here, for each
program, we considered a set of different queries and then we ran this
set with different number of threads. To do that, we implemented a
naive scheduler in Prolog code that initially launches the number of
threads required and then uses a \emph{mutex} to synchronize access to
the pool of queries. We experimented with a \textbf{Path} program
using a grid-like configuration and with two well-known ILP data-sets,
the \textbf{Carcino}genesis and \textbf{Muta}genesis data-sets. We
used the same 32 Core AMD machine, experimented with intervals of 8
threads up to 32 threads and the results that follow are the average
of 5 runs. Figure~\ref{fig_eval_tab_speedup} shows the average
execution time, in seconds, and the average speedup, compared against
the respective execution time with one thread, for running the naive
scheduler on top of these three programs with the \textbf{LFHT}
design.

The results show that our design has potential to speedup the
execution of tabled programs. For the \textbf{Path} benchmark, the
speedup increases up to 10.24 with 16 threads, but then it starts to
slow down. We believe that this behavior is related with the large
number of tabled dependencies in the program. For the \textbf{Carcino}
and \textbf{Muta} benchmarks, the speedup increases up to a value of
16.68 and 18.84 for 32 threads, respectively. Note that our goal with
these experiments was not to achieve maximum speedup because this
would require to take into account the peculiarities of each program
and eventually develop specialized schedulers for each one, which is
orthogonal to the focus of this work.

%%%%%%%%%%%%%%%%%%%%%%%%%%%%%%%%%%%%%%%%%%%%%%%%%%%%%%%%%%%%%%%%%%%%%%

\section{Conclusions}

We have presented a novel, simple and efficient lock-free design for
concurrent tabling. A key design decision in our approach is the
combination of hash tries with the use of a separate chaining closed
by the last node referencing back the hash level for the node. This
allows us to implement a clean design to solve hash collisions by
simply moving nodes between the levels. In our design, updates and
expansions of the hash levels are never done by using replacement of
data structures (i.e., create a new one to replace the old one), which
also avoids the need for memory recovery mechanisms. Another important
design decision which minimizes the bottlenecks and performance
problems leading to false sharing or cache memory side-effects, is the
insertion of nodes done at the end of the separate chain. This allows
for dispersing as much as possible the memory locations being updated
concurrently and, more importantly, reduces the updates for the memory
locations accessed more frequently, like the bucket entries for the
hash levels.

Experimental results in the context of Yap’s concurrent tabling
environment, showed that our design clearly achieved the best results
for the execution time, speedups and overhead ratios. In particular,
for worst case scenarios, our design clearly outperformed the previous
designs with a superb overhead always below 1.74 for 32 threads or
less. We thus argue that our design is the best proposal to support
concurrency in general purpose multithreaded tabling applications.

%%%%%%%%%%%%%%%%%%%%%%%%%%%%%%%%%%%%%%%%%%%%%%%%%%%%%%%%%%%%%%%%%%%%%%%%%%%%

\section*{Acknowledgments}

This work is partially funded by the ERDF (European Regional
Development Fund) through the COMPETE Programme and by FCT (Portuguese
Foundation for Science and Technology) within project SIBILA
(NORTE-07-0124-FEDER-000059). Miguel Areias is funded by the FCT grant
SFRH/BD/69673/2010.

%%%%%%%%%%%%%%%%%%%%%%%%%%%%%%%%%%%%%%%%%%%%%%%%%%%%%%%%%%%%%%%%%%%%%%

\bibliographystyle{acmtrans}
\bibliography{references}

%%%%%%%%%%%%%%%%%%%%%%%%%%%%%%%%%%%%%%%%%%%%%%%%%%%%%%%%%%%%%%%%%%%%%%

\end{document}